\documentclass[conference]{IEEEtran}
\IEEEoverridecommandlockouts
\usepackage{cite}
\usepackage{amsmath,amssymb,amsfonts}
\usepackage{algorithmic}
\usepackage{graphicx}
\usepackage{textcomp}
\usepackage{xcolor}
\def\BibTeX{{\rm B\kern-.05em{\sc i\kern-.025em b}\kern-.08em
    T\kern-.1667em\lower.7ex\hbox{E}\kern-.125emX}}

\begin{document}

\title{Self-Scaling Clusters and Reproducible Containers to Enable Scientific Computing
}

\author{\IEEEauthorblockN{Peter Z. Vaillancourt\IEEEauthorrefmark{1}, J. Eric Coulter\IEEEauthorrefmark{2}, 
Richard Knepper\IEEEauthorrefmark{3}, Brandon Barker\IEEEauthorrefmark{4}}
\IEEEauthorblockA{\IEEEauthorrefmark{1}\textit{Center for Advanced Computing} \\
Cornell University, Ithaca, New York, USA \\
Email: pzv2@cornell.edu}
\IEEEauthorblockA{\IEEEauthorrefmark{2}\textit{Cyberinfrastructure Integration Research Center} \\
Indiana University, Bloomington, IN, USA \\
Email: jecoulte@iu.edu}
\IEEEauthorblockA{\IEEEauthorrefmark{3}\textit{Center for Advanced Computing} \\
Cornell University, Ithaca, New York, USA \\
Email: rich.knepper@cornell.edu}
\IEEEauthorblockA{\IEEEauthorrefmark{4}\textit{Center for Advanced Computing} \\
Cornell University, Ithaca, New York, USA \\
Email: brandon.barker@cornell.edu}
}
\maketitle

\begin{abstract}
Container technologies such as Docker have become a crucial component of many software industry practices 
especially those pertaining to reproducibility and portability.  The containerization philosophy has influenced
the scientific computing community, which has begun to adopt -- and even develop -- container
technologies (such as Singularity).  Leveraging containers for scientific software often
poses challenges distinct from those encountered in industry, and requires different
methodologies.  This is especially true for HPC.  With an increasing number of options for 
HPC in the cloud (including NSF-funded cloud projects), there is strong motivation
to seek solutions that provide flexibility to develop and deploy scientific software 
on a variety of computational infrastructures in a portable and reproducible way.  
The flexibility offered by cloud services enables virtual HPC clusters that scale on-demand, and the
Cyberinfrastructure Resource Integration team in the XSEDE project has developed a set
of tools which provides scalable infrastructure in the cloud.
We now present a solution which uses the Nix package manager in an MPI-capable Docker container
that is converted to Singularity. It provides consistent installations, dependencies, and environments
in each image that are reproducible and portable across scientific computing infrastructures.
We demonstrate the utility of these containers with cluster benchmark runs in a 
self-scaling virtual cluster using the Slurm scheduler deployed in the Jetstream
and Aristotle Red Cloud OpenStack clouds.  We conclude this technique is useful
as a template for scientific software application containers to be used in the
XSEDE compute environment, other Singularity HPC environments, and cloud computing environments.
\end{abstract} 

\begin{IEEEkeywords}
Containers, Docker, Singularity, Slurm, Cluster, Cloud, Scientific Computing, HPC, MPI, Cyberinfrastructure
\end{IEEEkeywords}

\section{Introduction}
\label{sec:intro}
The Cyberinfrastructure Resource Integration (CRI) group of  
Extreme Science and Engineering Development (XSEDE) \cite{towns2014xsede} provides software and services to bring
best practices from the National Supercomputing Centers to university campuses
interested in implementing their own computational infrastructure. By
providing software and documentation for campus adopters, CRI has been able to
assist in the creation of cyberinfrastructure at a number of 
campuses, enabling them to provide computational resources to their faculty
members\cite{XCBC-LessonsLearned}.  As more organizations utilize cloud resources in order to provide
flexible infrastructure for research purposes, CRI has begun providing tools to
harness the scalability and utility of cloud infrastructure,
working together with the members of the Aristotle Cloud Federation Project\cite{knepper2019red}.
Red Cloud, the Cornell component of the Aristotle Cloud Federation, provides an
OpenStack cloud services much like the Jetstream resource, that is available
both to Cornell faculty as well as to Aristotle members.

CRI focuses its activities on knitting together cyberinfrastructure 
resources at multiple institutions.  While the XSEDE Federation presents
the resources of sundry service providers at different levels of interoperability \cite{xsedefederation}, 
there are many campus cyberinfrastructure installations
that are focused on meeting local needs.  Expertise in cluster administration
is a scarce resource, and not all institutions that elect to provide
a centralized computing cluster are immediately able to leverage
technical and policy knowledge to implement and maintain it. The CRI team
assists these campuses in set-up, maintenance,
and policy generation. If the institution has interest, CRI assists with integration
into the larger XSEDE framework. CRI has worked with 6 different institutions
in 2019 and 2020 to provide support for cyberinfrastructure installations,
and has directly assisted in bringing over 1 PetaFLOP of computing capacity online,
in addition to advising on other cyberinfrastructure activities.

Previously, CRI introduced the virtual cluster toolkit \cite{coulter2017programmable,crijetstreamcluster}, which allows for the creation of
cluster computing resources on cloud resources.
It was specifically created for the Jetstream \cite{Jetstream}
research cloud, but is adaptable to other cloud resources, as seen in our deployment 
to Red Cloud in this paper.  This virtual cluster toolkit
provides infrastucture that can be used for a variety of jobs including
multi-node jobs, managed by the Slurm scheduler \cite{SLURM}.  

CRI has also taken efforts to expand
software offerings which take advantage of the benefits of cloud computing
models: reproducible application containers and self-scaling clusters built
on cloud infrastructure.  Reproducible containers take advantage of 
container runtime environments which allow applications to run in
lightweight virtual spaces that make use of the underlying operating system,
while providing reliable installation of a set of software and dependencies
at a given version.  Self-scaling clusters rely on common cluster system images
and the Slurm scheduling tool in order to create and destroy additional virtual
machines as jobs are placed into the Slurm queue. Reproducible containers
provide reliable, portable applications with the same constituent software
at every instantiation, while self-scaling clusters provide efficient use of 
computational infrastructure without incurring more usage than the 
applications require.

\section{Compute Environment Choices}
\label{sec:ce-choices}
\subsection{XSEDE Compute Environment}
\label{sec:xsede-ce}
Researchers at U.S. universities and federally-funded research and development centers (FFRDC's) consistently report
a lack of sufficient computational capability available to them \cite{stewart-technical-2011,ACCI-Taskforce}.  The XSEDE
project provides a socio-technical system which facilitates access to resources provided through
the XSEDE federation, but despite the considerable resources available,
not all requests can be accommodated.  XSEDE allocation proposals outstrip allocable resources by
a factor of three.  By passing on best practices and interoperability to other institutions,
the XSEDE project can foster the extension of computational capacity beyond what is available
through the project alone, and in addition, can make it easier for researchers to
move between systems.  By providing common filesystem layouts, software, and scheduling
systems, adopters of XSEDE software capabilities can make it easier for researchers to
work on XSEDE systems as well as other systems which have adopted similar affordances,
making it easier to move between resources and facilitating scaling up or down in 
compute capability as necessary.

However, not all XSEDE resources and not all campus resources can be exactly the same.
Local context dictates particular choices which affect individual system 
implementations as well as which software is made readily available.
Analytical workflows cannot be instantaneously replicated on different systems,
and barriers to computational transparency between systems will always be a factor.  For XSEDE
systems, not only are there dedicated, system-specific allocations which determine access to resources
through the XSEDE access system, but system differences and separation of file systems mean
that researchers must make changes to their workflows in order to move to different
systems within the XSEDE framework. Researchers making use of the national infrastructure
at a variety of sites, facing switching costs to move between systems, require a common set of tools 
which facilitate their use of computational resources. 

\subsubsection{XCBC}
\label{sec:xcbc}
In an effort to help solve the problems facing campus resource providers
in giving their researchers access to quality High Performance Computing (HPC) 
environments, the XSEDE CRI team developed a toolkit called the
XSEDE-Compatible Basic Cluster (XCBC) to easily build a local HPC system
conforming to the basic environment available on most XSEDE resources, 
modulo differences in specialized hardware or site-specific software.
This project evolved\cite{CB:XCBC-paper} from being based on the Rocks\cite{Rocks} cluster
management software to the OpenHPC\cite{OpenHPC-paper} project with 
Ansible\cite{Ansible}. The CRI team has performed a dozen
in-person site visits to help implement HPC systems using this 
toolkit, in addition to dozens more remote consultations. Discussions
with sites implementing the toolkit\cite{CLangin-PEARC19,Bentley-PEARC19,CBPanel-PEARC19} have shown that the local XSEDE-like
environment does in fact help researchers onboard to XSEDE more easily, 
in addition to providing an environment with low wait times compared
to the full queues on larger systems, where researchers may wait
for days before a particular job runs.

\subsubsection{Singularity in HPC}
\label{sec:sing-in-hpc}
The problem remains, however, that individual researchers may not have
the software they need available on either a national or a local campus system,
and getting new custom software installed is non-trivial for the 
administrators of those systems. In many cases, this can be addressed through the
use of containers, which have sparked a revolution in horizontal scaling
of software in the enterprise world. In addition, containers provide the
opportunity to utilize differing versions of software than those installed,
or to package private or custom software for development purposes.

There are significant challenges to using containers effectively on HPC systems, 
however, due to security concerns, complexity of installations for scientific software,
and lack of widespread adoption as a distribution method for scientific software.
The popularity and prevelance of Docker\cite{docker}
outside of the HPC community notwithstanding, security concerns around the
fact that a user running docker has root access to a system prevent
its use on HPC systems without modifications to mitigate 
risk.  Newer container runtime environments have been developed with an eye towards 
solving these security concerns and even designed specifically for HPC systems, 
such as CharlieCloud\cite{charliecloud} and Singularity\cite{singularitypaper,singularity}.

Singularity has become the de-facto preference for containers in many HPC
environments, including the XSEDE compute environment.  Singularity
is available on all XSEDE allocated resources, and included in the XCBC toolkit as
part of OpenHPC, thus becoming available on many campus systems.
Other HPC systems using versions of OpenHPC, even outside of the XCBC toolkit,
can also trivially provide access to Singularity to their users.  This prevalence within HPC
makes Singularity an attractive option for the distribution of scientific software and
dependencies, which is further helped by Singularity support
for the conversion of Docker container images to Singularity\cite{singularitydocker}.
So long as best practices for conversion to Singularity are followed (or checked in the 
case of pre-existing images), the same Docker container that was developed for deployment
to a cloud environment can be converted to Singularity for effective use on an HPC resource.
An essential component of both Docker and Singularity containers is that the 
common means for instantiation relies on a pull method, which generally grabs the 
most recent version of the container image. 
This results in
the situation that execution of the ``same" container at different times could use different
versions of the included software, potentially affecting the results of computation.  This 
complicates the replication and reproducibility of calculations later on, and necessitates 
a replication-friendly container architecture.
\subsection{The Need for Cloud-Integrated Computational Software}
\label{sec:cloud-motivation}

In many cases, researchers do not have access to local HPC hardware on
which to test and profile their software, before moving to large-scale XSEDE environments such as Comet or
Stampede2. In order to provide an HPC testbed environment to such
researchers, the XCRI team has developed a self-scaling virtual 
cluster\cite{HARC19-VC} for use
on Openstack clouds, such as Jetstream\cite{Jetstream}. This toolkit
provides researchers with a basic HPC system, based on the 
OpenHPC\cite{OpenHPC-paper} project
using the same software as those powering the systems available 
within XSEDE. This allows for rapid prototyping, scaling tests, and scientific 
software development within an XSEDE-like environment, without the associated ``cost''
encountered on large systems
of long wait times in filled queues, environmental roadblocks, having to ask overwhelmed
sysadmins to help with your software install, and navigating policies that favor large 
highly-tuned jobs over small productivity-oriented jobs.

Nonetheless, over the course of providing virtual HPC environments for dozens of projects
\cite{HARC19-VC}, 
it has become quite clear that the administrative overhead of maintaining 
scientific software does not scale well within a small team, \emph{nor} does it
effectively empower users to easily transfer software and workflows into XSEDE
without assistance.
While the premise of familiarizing users with an HPC environment holds in situations 
where researchers are new to HPC, things fall apart in the case of established 
research teams who want to take customized software from a cloud environment with 
root access into a heavily managed multi-user system. Thus, we have begun 
transitioning from custom software installations on a per-cluster basis to 
helping users containerize their software for use in multiple XSEDE environments.
By providing container templates based on best practices for the container runtimes available
on national resources, we are able to further erode the barriers to computing that 
prevent researchers from making maximally effective use of their time and allocations.
To further reduce barriers to scientific computing, we provide Docker container
templates that can run in a variety of cloud environments -- while still maintaining
best practices for conversion to Singularity -- to enable consistency of
software and environment across different cyberinfrastructures.  Thus, researchers
can employ these container templates for greater access to computing through 
whatever means (i.e. cloud or HPC resources) are available to them.

\section{Reproducible Application Containers}
\label{sec:repro-app-containers}
When developing containers for scientific software applications, 
it quickly becomes aparent that subtle version changes, differing compilation or configuration steps, or any
number of other nuances of the build and installation process can cause unexpected behavior or
even failure to build.  Though version pinning is the best practice for software installation
inside a container in order to achieve consistency, it is not common practice.  Additionally, scientific
software in some domains -- especially legacy codes used in tandem with newer packages -- often has
an extensive list of dependencies, making version pinning a tedious task for researchers
managing their own development code, environment, and dependencies.  Hence, even many currently
available container specifications fail to build when attempted a short time period after they
were written.  In spite of this discrepancy, if the container images are publicly available, they
can continue to be used for much longer than the build specification so long as no changes are needed
in version or other configurations.  On the other hand, scientific software development often
involves an iterative process, where compilation steps and configuration choices change over time.
If other users or even the original user return to the container build specification to make
changes, they can be thrust back into the role of a system administrator tracking down the source
of discrepancies in software installation and setup as compared to previous instantiations of
the container image.  Moreover, the conversion of a container between the Docker and Singularity container runtime
environments does not guarantee consistency, or reproducibility, of software installations or
configurations even when best practices are followed.  

Reproducibility and associated terms like replicability and reliability represent considerable
issues for researchers in demonstrating that their research is supported by data that is correctly analyzed, 
with results that are descriptive of the subject of inquiry, with consistent support for the
inferences that those results yield.  These components have been described as ''methods reproducibility'', 
''results reproducibility'', and ''inferential reproducibility'' \cite{goodman2016does}.
Funding agencies such as the NSF and NIH state that the ability to reproduce research is the
basis of research findings that are ''believable and informative'' \cite{bollen2015reliable,collins2014policy}.
With researchers who want to provide a means for reproducing their analyses facing
a considerable number of variables, the current state of scientific software in containers
represents an issue in ''methods reproducibility'', which we believe can be addressed by
providing a framework that delivers software components of the same version from the original
analysis at every build and instantiation time, providing transparency into the components used, and facilitating
the simple reproduction of those software components by later scientists even after many years. 
In order to support this framework, we leverage the Nix package manager and expression language.

\begin{figure}
    \centering
    \includegraphics[width=\columnwidth]{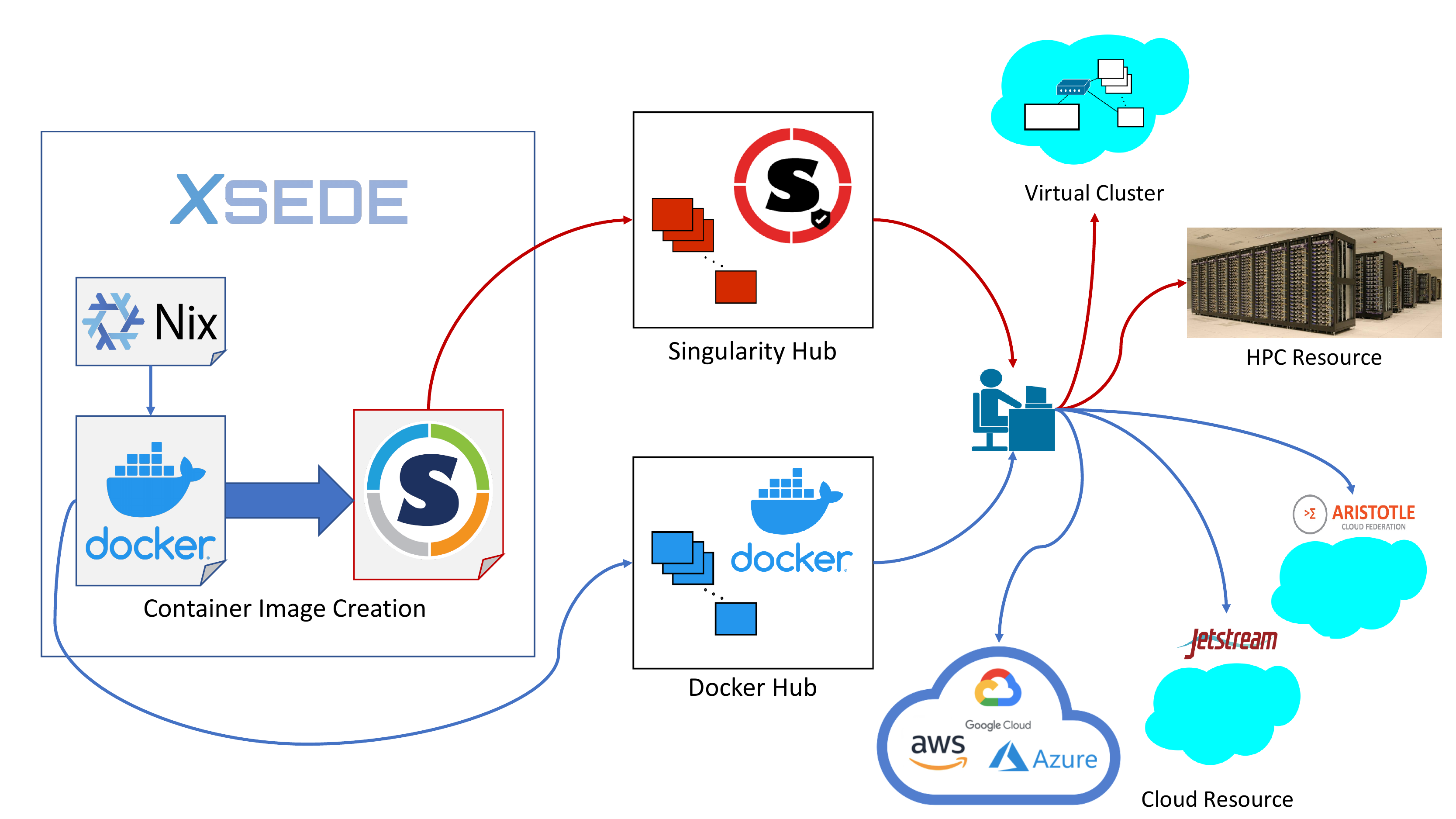}
    \caption{A depiction of the process from container template development by the XSEDE XCRI team
    through deployment of Singularity or Docker by the user in various environments.
    }
    \label{fig:ctl}
\end{figure}

\subsection{NIX}
\label{sec:nix}

Nix is a package manager and associated language (Nix Expression Language\cite{nix-expression}) 
for specifying package builds and dependencies \cite{nixthesis, nix}.  It has two major sister 
projects, the nixpkgs package collection that contains many thousands of package definitions 
in a single GitHub repository\cite{nixpkgs}, and the Linux distribution NixOS\cite{NixOS}, which 
uses the former projects on top of the Linux kernel and systemd to realize the vision of an
entire Operarting System (OS) environment that is easily reproducible. This reproducibilty is achieved as part of
the Nix-language architecture and by pinning the particular version (or versions) of
nixpkgs being used. The Aristotle Science Team has taken advantage of this feature of Nix
within Docker containers for cloud deployments of scientific applications\cite{PEARC20-paper,nixtemplates,lp-repo,waterpaths}, 
building upon previous work which demonstrated it's use in scientific computing
and HPC\cite{nixhpc,nix-reproduce}, and here we apply the usage and extend it to Singularity containers.

Although most are familiar with the idea of pinning versions in various settings to obtain 
reproducibility, the Nix language takes this a step further.
Package specifications  -- called Nix \textit{derivations} -- are just a kind of
Nix function written in the Nix Expression Language. The function takes inputs such as
other prerequisite packages (i.e. compilers), build configurations for the same package,
possibly that package's dependencies, and any source or binary files needed.  It is important
to note that this sort of specification is always more precise than just specifying a package's
version. The entire installation process can be customized within a Nix expression, if desired, 
included everything from simple directory and file management to complicated or intricate
steps for compilation or configuration.  Even beyond this, the Nix language is a pure functional
language, meaning that one will always get the same output for a given input.  The inputs
are hashed, whether they other nix expressions, binary or source blobs downloaded over the
internet, or configuration files that are read in by the Nix derivation.  If a hash changes
in the build, a new package is created with the new hash recorded.  For instance, by
changing a build setting, one would wind up with two installations of the same package
indexed by the hash of the derivation output.  Applications requiring the package as
a dependency will automatically get the correct package loaded in the environment
depending on what configuration settings they specify for the package in their derivations.


While using Nix in containers is not the most widepsread use of the Nix package manager,
it can be used in Windows Subsystem for Linux (WSL), Mac OS X, and practically any Linux
distribution, including in Linux containers.  By using a container that includes the Nix
package manager, we can create distribution-independent and largely environment-independent
solutions that are portable across all environments supporting Nix (whether inside a container
or not).  One caveat to this that we have encountered is with MPI applications, where the
implementation and version of MPI within the container must match that of MPI on the host
(within a range of major release version).  This is because for distributed MPI jobs, there
are components of MPI that must run at a privileged level on the host, e.g. when dealing
with network fabric.  To migitate this challenge, our container templates will be 
flexible with respect to MPI versions and implementations where possible.

\subsection{Container Template Library}
\label{sec:library}
As described above,
the CRI team has begun to curate a Container Template Library (CTL) to 
share with the wider scientific computing community.  The goal of the project is to create
easy-to-use, and reproducible templates for containerizing scientific
software using Docker or Singularity with Nix.  A ``template'' consists of an open source GitHub
repository with the Dockerfile (build specification for Docker) and any associated scripts
(such as Nix expressions) used to build the container, the Docker image hosted on Docker
Hub\cite{dockerhub}, an open source GitHub repository with the Singularity definition file
(build specification for Singularity) and any example Slurm scripts, and a 
Singularity image hosted on SingularityHub\cite{singularityhub}.  
The benefit
of providing \textit{templates} for containers, as opposed to just the container images,
lies in empowering the user to be
able to modify the build to suit their needs,
knowing that it will be reproducible and result in a consistent environment.

Fig.~\ref{fig:ctl} demonstrates the entire process for a single container template.  The
CRI team begins by writing any needed Nix expressions (where standard Nix packages cannot
be used) and providing a Dockerfile that properly builds and installs the environment, packages,
dependencies, etc.  These are used to produce a public Docker image that is hosted on Docker
Hub, which the user can then pull to deploy on any cloud resource they have access to.
Alternatively, the user can return to the Dockerfile or Nix expression to make modifications
or add their own software, and then build their own image.  The CRI team also creates a
Singularity definition file which exercises the built-in Singularity methods to convert
a Docker image to a Singularity image, and adds tests and scripts to run the container.
The definition file is then added to the SingularityHub collection and built to create a
public image which the user can pull to either a virtual cluster in
the cloud or any HPC resource they have access to.  Current available containers and images include:
\begin{itemize}
    \item Docker base container with Nix\cite{docker-nix-base,dhub-nix-base}
    \item Singularity base container with Nix\cite{singularity-nix-base,shub-nix-base}
    \item Docker container with Nix and OpenMPI\cite{docker-nix-ompi,dhub-nix-ompi}
    \item Singularity container with Nix and OpenMPI\cite{singularity-nix-ompi,shub-nix-ompi}
\end{itemize}
In this paper, we have chosen the Singularity container with Nix and OpenMPI
to deploy and run on a virtual cluster.

\section{Virtual Clusters}
\label{sec:VCs}
As discussed in Section \ref{sec:cloud-motivation}, the XSEDE-like Virtual Cluster (VC)
has been developed to provide researchers with an environment for testing and scaling
scientific workflows. The basic environment provides the following features, which 
are often the hardest for a researcher to navigate upon gaining access to a new 
system:
\begin{itemize}
    \item \textbf{Slurm} - The scheduler and resource manager used to split the work of multiple users across multiple nodes
    \item \textbf{Lmod} - The environment module system in use across XSEDE resources, allowing for easy setting of environment variables related to specific software packages\cite{layton2015lmod}
    \item \textbf{Singularity} - The container runtime environment available across XSEDE HPC resources, designed specifically to ease use of MPI codes and integrate with Slurm.
    \item \textbf{Shared storage} - every HPC system has different requirements for shared storage and scratch space; the VC uses a simple system of three shared spaces (user home directories, a working directory space, and a shared space for installed software).
\end{itemize}

These virtual HPC environments have also proven useful in educating users and 
potential sysadmins about what it takes to run an HPC system. They have been used in a series of tutorials\cite{JOCSE-JSTut,coulter2017programmable},
including
in-person workshops at George Mason University and Clark Atlanta University.

\subsection{Virtual Cluster Architecture}
\label{sec:vc-arch}
\begin{figure}
    \centering
    \includegraphics[width=\columnwidth]{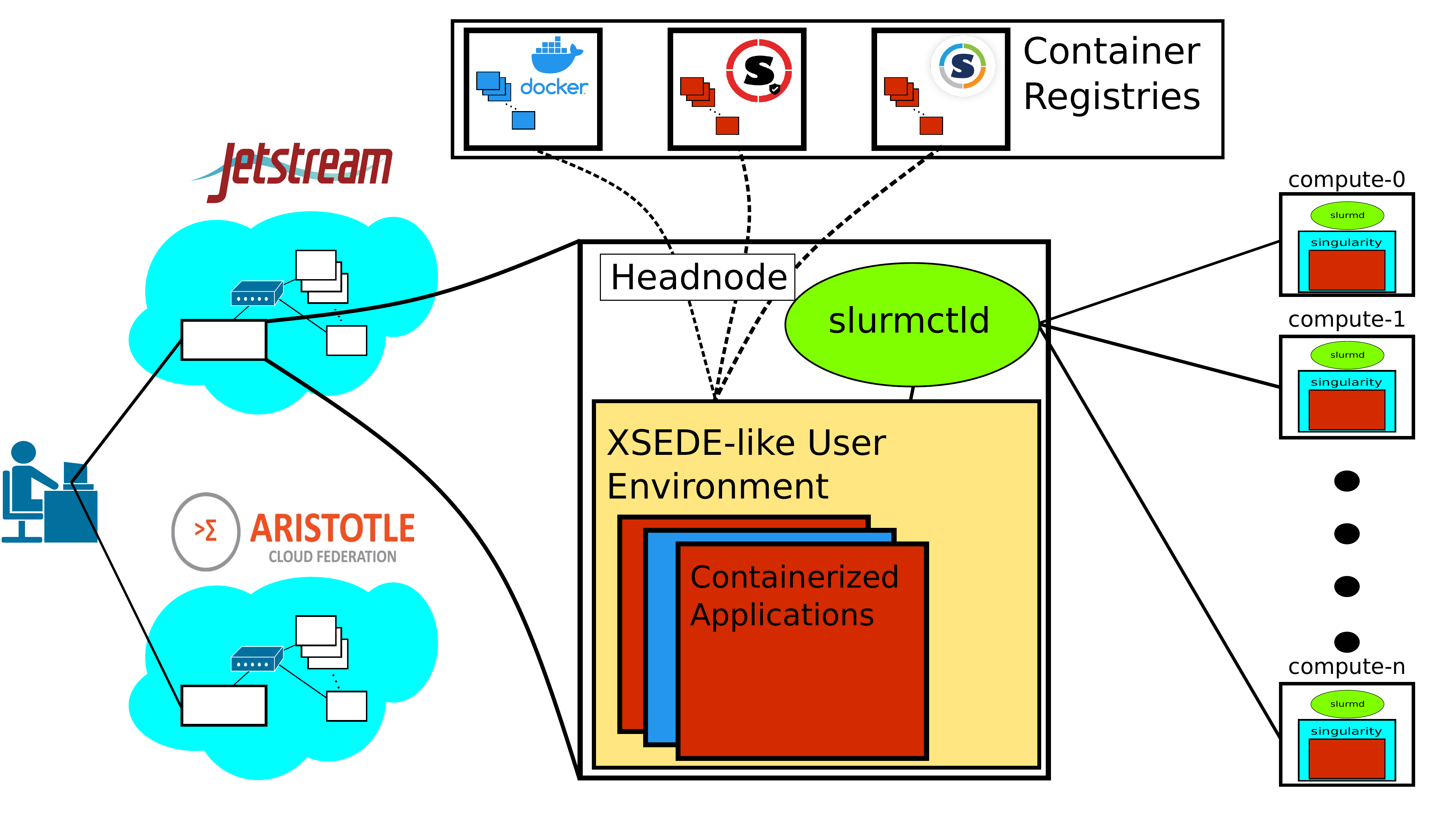}
    \caption{Overview of the VC architecture from the user
    perspective.}
    \label{fig:vc_overview}
\end{figure}

The VC infrastructure is primarily centered on the headnode, which also
provides users login access (and an optional graphical environment). In a
sense, the VC consists solely of the headnode - all work is done on ephemeral instances, 
created when a user submits a job to Slurm, and destroyed when there is no more
work in the queue. As we show in Fig.~\ref{fig:vc_overview}, from a user perspective,
it is possible to deploy HPC-style infrastructure across clouds, and  run jobs using 
reproducible containerized software. The VC manages compute (worker) instances without user 
intervention, allowing for efficient usage of resources. The container images are 
easily made available to other researchers via container registries, allowing for 
easier collaboration and greater transparency of published work.

\section{Deployment Process}
\label{sec:vc-deployment}
The current VC can be deployed via two commands after
cloning the GitHub repository\cite{crijetstreamcluster} hosting the code.
The first creates the headnode which will: spawn worker nodes via
Slurm, contain the user environment and software (container 
images), and provide the filesystem shared to worker nodes. The
second command installs all of the required software on
that node.

For production useage, edits must be made to two files to reflect the size
of jobs expected on the VC, to increase the size and number of
worker nodes, or create larger shared storage volumes. Otherwise,
the addition of scientific software can be done simply by
pulling down a Singularity image, and
jobs can be run without further intervention. A more detailed 
discussion of the configuration of any individual VC is provided in \cite{HARC19-VC}. 
For clarity in the following discussion in Section \ref{sec:mpi-app-run}, 
it is worth mentioning that the software on the cluster is 
installed from repositories provided 
by the OpenHPC\cite{OpenHPC-paper} project, including Singularity version 3.4.1 and 
OpenMPI version 3.1.0 at the time of writing.

\subsubsection{Mutation for Different Clouds}
\label{sec:vc-mutation}
In a similar vein to the issues discussed in Section \ref{sec:xsede-ce},
there still remain barriers to implementing the same 
infrastructure on different research cloud providers \emph{even} in the case where 
the underlying software is the same. For the purposes of
this paper, we created example clusters on Jetstream
and Red Cloud, to demonstrate reproducibility of
the entire stack and workflow on different cloud environments,
which still requires modification of the scripts used
to create the VC infrastructure. This problem is also commonly
encountered on public clouds as well, given the rapid
pace of changes to APIs and charging mechanisms. 
The differences in cloud providers using OpenStack primarily  
stem from the vast array of underlying services that make
up a cloud system, which naturally leads to different 
network configurations, instance flavors, etc. across
providers. 

Specifically, in order to re-create the VC architecture on 
Red Cloud rather than Jetstream, there were three changes to be
made:
\begin{enumerate}
    \item \textbf{Change base image name for instances}:
    Red Cloud uses a simpler image naming scheme 
    than the more detailed convention used on Jetstream, 
    where there is a larger user base and set of images.
    \item \textbf{Change instance flavor names}:
    Red Cloud uses AWS-style names (c1.m8, c2.m16, etc.) for instance ``flavors"
    as opposed to Jetstream's OpenStack-default convention (m1.tiny, m1.quad, etc.).
    \item \textbf{Update network creation scripts}:
    Red Cloud requires explicitly setting the internal DHCP servers when creating a private network, where Jetstream sets this by default, which can lead to confusion when changing the network configuration of extant instances.
\end{enumerate}

For the user deploying a VC in Red Cloud, these
changes amount to small edits in four files used to control the
VC configuration, which will be simplified in future releases.

\subsection{MPI Application Run}
\label{sec:mpi-app-run}
As an example application, we have chosen the ever-popular 
HPL benchmark\cite{HPL}, the measure of choice for the Top500 list, in
which HPC systems are ranked by Floating-point Operations Per Second
(FLOPs). 
The HPL software is included in the Singularity container with Nix
and OpenMPI discussed in Section \ref{sec:library}.
Tuning the HPL benchmark for maximum performance
is a highly involved process for large systems\cite{hpl-paper}. 
In light of this, we do not present these results 
as true measures of the performance of the VC, but as simple 
representation of the fact that our container builds allow us to run
MPI codes across multiple systems with very little effort. The main
effort in running these jobs, in fact, was in creation of the 
input file for the HPL run, using the tool provided by Advanced 
Clustering\cite{ac-hpl-tuner}, which nicely codifies the advice provided by the 
creators of HPL\cite{netlib-HPL}. It is worth noting again that the
version of OpenMPI provided by nixpgs is 4.0.1, which ``just worked" 
in conjunction with the host OpenMPI at version 3.1.0. These jobs
were run using the ``hybrid model"\cite{singularity-readthedocs} of MPI execution
in which the host version of MPI is used to execute the singularity command.
For the full Slurm script, see \cite{singularity-nix-ompi}, but the main command
looks like the following:
\begin{verbatim}

mpirun -np $NUMPROCS singularity \ 
       exec hpl.sif xhpl ./HPL.dat

\end{verbatim}
This does require that the MPI version internal to the container is
compatible with that of the host, but as demonstrated here, that
covers a wide range of versions, leading to maximum flexibility. Of 
course, we have also designed our Dockerfiles to be flexible with 
respect to the internal version and implementation of MPI used.

\begin{table}
    \centering
    \begin{tabular}{c||c|c||c|c}
         System     & 1 Node GFLOPS & Rmax  & 4 Node GFLOPS & Rmax \\ \hline
         Jetstream  &  146          & 160   & 500           & 640 \\ \hline
         Red Cloud  &  105          & 140   & 412           & 540 \\ \hline
    \end{tabular}
    \caption{Results from best of three runs of HPL in Singularity containers on two Openstack Clouds.}
    \label{tab:hpl-results}
\end{table}

As shown in Table \ref{tab:hpl-results}, the results generally 
indicate performance in the range of $73-78\%$ of the theoretical
maximum performance of the system, which in the authors' experience 
running naive builds of HPL on bare metal is not bad at all - it 
is fairly likely that with concerted effort, near bare-metal performance
could be extracted, based on other work comparing performance losses
due to both containerization and virtualization\cite{perf1,perf2,perf3,perf4,perf5}.
For example, during the initial testing of the Jetstream platform, virtualization
induced only a 3\% hit in HPL performance\cite{Jetstream}. This is not
to say that it is without dangers, as virtualization can have differing
impacts depending on the actual workload\cite{js-virtualization-impact}.

\section{Practical Outcomes}
\label{sec:practical-outcomes}
Practically, this work demonstrates a method for creation of 
infrastructure-independent (within the bounds of container runtime 
environment and MPI implementations and versions), reproducible containers
for scientific software, with or without the requirements of running
across multiple nodes. While we focus predominantly on software requiring
MPI, this is only to illustrate the use of the Nix plus Docker/Singularity
stack for the more difficult case where multiple nodes are required. 
Scientific software that would not previously be, can now become portable
and reproducible with the flexibility develop and deploy on a personal computer,
cloud computing environment, or an HPC resource.  Furthermore, the
user can enjoy consistency of environment across deployments, and even
customize the environment to serve their research projects.
We have additionally demonstrated a simple, flexible HPC-style 
infrastructure which is deployable across multiple cloud providers
at nearly the push of a button, which can provide an invaluable
testing or bursting environment for researchers lacking in local 
resources.  Additionally, this infrastructure is already battle-tested
behind the scenes of numerous Science Gateways\cite{HARC19-VC,Interactwel-PEARC19,DR-PEARC19,CCGrid2017-VC}.

\section{Conclusion} 
\label{sec:conclusion}
In terms of future work, a few clear points for improvement are 
methods to allow expansion of the VC infrastructure to public clouds,
possibly through the use of Terraform \cite{terraform}. While some providers 
offer services that putatively offer HPC-style computation, the cost
in terms of knowledge-gathering is often quite high, and is not even
by default elastic, exposing the user to the potential for massive
costs if worker nodes are not fully deleted subsequent to jobs finishing.
In the same vein, a few small tweaks to the current VC configuration 
methods should allow users to specify the size of the desired cluster
either through editing a single file, or providing a flag at build-time.

For the CTL, the next step for this team is to provide a larger set of
stable templates for 
reproducible containers, based on Nix and Docker and converted to Singularity. 
We plan to also include in-depth documentation for the customization of the
environment, adding scientific applications to the containers, and more
example Slurm scripts.  The ultimate goal of the project is a library
containing a variety of scientific applications that have already been
built within this framework, which the community can employ to enable
the rapid deployment of their codes.
This library of containers will be tested across the nationally available 
cyberinfrastructure for usability, and will also be leveraged by Science Gateways projects,
in particular those powered by Apache Airavata, in order to expand
the computational resources available through easy-to-use web 
interfaces, much as the VC has done for expanding the use of 
the Jetstream resource\cite{HARC19-VC}.

\bibliographystyle{IEEEtran}
\bibliography{my,barker,pete,rich}
\end{document}